\documentclass[nobibnotes,preprintnumbers,nofootinbib]{revtex4}
\usepackage{epsfig}
\begin{document}

\title{\mbox{}\\[10pt]
 Probing lepton flavor violation with nonzero $U_{e3}$ and leptogenesis\\
  through $A_4$ symmetry breaking}

\author{Y. H. Ahn
\footnote{Email:yhahn@phys.sinica.edu.tw}}

\affiliation{Institute of Physics, Academia Sinica, Taipei 115, Taiwan}

\date{\today}

\begin{abstract}
We study how lepton flavor violation (LFV) is related to the reactor angle $\theta_{13}$, LHC/dark matter (DM) signal with a successful leptogenesis in a radiative seesaw model through $A_4$ flavor symmetry breaking, in the light of the deviation of tri-bimaximal mixing (TBM) angles and the recent precision oscillation data. To do this, we consider the normal mass spectrum of light neutrino giving a successful leptogenesis at electroweak (EW) scale or more with a relatively large $\sin\theta_{13}\simeq{\cal O}(0.1)$, where a successful leptogenesis requires LHC/DM signal $m_{-}\lesssim$ a few keV in our scenario, and we show that the measurement of $\tau\rightarrow\mu\gamma$ is strongly dependent on $m_{-}$ and the leptogenesis scale. Especially, we show that $\mu\rightarrow e\gamma$ and $\mu-e$ conversion are proportional to the value of $\theta_{13}$, and the measurements of both $\mu\rightarrow e\gamma$ (and/or $\mu-e$ conversion) and $\theta_{13}$ can give a prediction of $\tau\rightarrow\mu\gamma$ with a successful leptogenesis.

\end{abstract}

\maketitle %
\section{Introduction}

The observed neutrino oscillations and baryon asymmetry of the Universe (BAU) as well as the existence of DM clearly imply physics beyond the standard model (SM). One of the most striking developments in particle physics beyond the SM is the experimental establishment of neutrino data shown in Table-\ref{tab:data};
\begin{widetext}
\begin{table}[th]
\begin{center}
\begin{tabular}{|c|c|c|c|c|c|} \hline
 & $\Delta m^{2}_{\rm Sol}/10^{-5}\mathrm{\ eV}^2$ & $\sin^2\theta_{12}$ & $|U_{e3}|$ & $\sin^2\theta_{23}$ &
$\Delta m^{2}_{\rm Atm}/10^{-3}\mathrm{\ eV}^2$ \\ \hline \hline
Best-fit        &     7.67     &  0.312          &  0.126          &
0.466          &  2.39 \\ \hline
$1\sigma$ & $7.48 - 7.83$ & $0.294 - 0.331$  & $0.077 -
0.161$  & $0.408 - 0.539$  & $2.31 - 2.50$ \\ \hline
$3\sigma$ & $7.14 - 8.19$ & $0.263 - 0.375$  & $<0.214$
& $0.331 - 0.644$  & $2.06 - 2.81$ \\ \hline
\end{tabular}
\caption{\label{tab:data}Current best-fit values as well as 1 and $3\sigma$ ranges of the oscillation parameters \cite{bari}.}
\end{center}
\end{table}
\end{widetext}
Although neutrinos have gradually revealed their properties in various experiments since the historic Super-Kamiokande confirmation of neutrino oscillations \cite{Fukuda:1998mi}, properties related to the leptonic CP violation are completely unknown yet. In addition, the large mixing values of $\theta_{\rm sol}\equiv\theta_{12}$ and $\theta_{\rm atm}\equiv\theta_{23}$ may be telling us about some new symmetries of leptons that are not present in the quark sector and may provide a clue of the nature among quark-lepton physics beyond the SM. Recently, there have been some attempts to explain the mass and mixing pattern in the leptonic sector, which is the most popular discrete $\mu-\tau$ symmetry~\cite{mutau}. Nevertheless, E.Ma and G.Rajasekaran~\cite{Ma:2001dn} have introduced for the first time the $A_{4}$ symmetry to avoid mass degeneracy of $\mu$ and $\tau$ under $\mu-\tau$ symmetry.
In models of $A_4$ symmetry~\cite{A4}, the so-called TBM pattern comes out in a natural way, $\sin^{2}\theta_{12}=1/3, \sin^{2}\theta_{23}=1/2$ and $\sin^{2}\theta_{13}=0$ which is fully compatible with the present knowledge of neutrino oscillation data in Table-\ref{tab:data} at $3\sigma$. In the light of the CP violation from the neutrino
oscillations, the TBM indicates that the CP asymmetry $P(\nu_{\mu}-\nu_{e})-P(\bar{\nu}_{\mu}-\bar{\nu}_{e})$ is vanishing. Therefore, finding non-vanishing but small mixing element $U_{e3}$ would be very interesting in the sense that the element is closely related to leptonic CP violation~\cite{Barr:2000ka}. Moreover, the recent analysis based on global fits of the available data gives us hints for $\theta_{13}>0$ at $1\sigma$; as the best-fit value $\sin\theta_{13}\simeq{\cal O}(0.1)$\cite{Fogli:2009ce,nudata}. So, the precise measurement of $\theta_{13}$ is a crucial test of the models.

Besides the mystery of the mixing pattern, tiny neutrino mass is one of the most challenging problem beyond SM. Recently, E.Ma introduced the so-called radiative seesaw mechanism \cite{Ma:2006km} where the neutrino masses are generated through one-loop mediated by a new Higgs doublet and right-handed neutrinos obeying an additional $Z_2$ symmetry: a $Z_{2}$-odd quantum number is assigned to a leptonic Higgs doublet $\eta=(\eta^{+},\eta^{0})$ and three right-handed singlet fermions $N_{i}$ while all the SM particles are $Z_{2}$-even. After electroweak symmetry breaking, the $Z_{2}$ symmetry is exactly conserved and $\eta$ will not develop a VEV, that is $\langle\eta^{0}\rangle=0$, while the standard Higgs boson get a VEV, which means the Yukawa coupling corresponding to $Z_{2}$-odd Higgs doublet will not generate the Dirac mass terms in neutrino sector. Thus, the usual seesaw mechanism does not work any more and we naturally have a good candidate of DM corresponding to the lightest $Z_{2}$-odd particle or Large Hadron Collider (LHC) signals through the standard gauge interactions in our model. In addition, since the existence of the flavor neutrino mixing for the three neutrinos $\nu_e,\nu_\mu,\nu_\tau$ implies that the individual lepton charges, $L_\alpha,(\alpha=e,\mu,\tau)$, are not conserved~\cite{Bilenky:1987ty}, the observation of neutrino oscillation have the possibility of measurable branching ratio for charged lepton LFV decays such as $\mu\rightarrow e\gamma,\tau\rightarrow e\gamma$ and $\tau\rightarrow\mu\gamma$, etc.. Experimental discovery of such lepton rare decay processes is one of smoking gun signals of physics beyond the SM.
Besides, the observed BAU can be explained by the mechanism of leptogenesis\cite{Fukugita:1986hr,Langacker:1986rj}.
If this BAU originated from leptogenesis, then CP asymmetry in the leptonic sector must be broken. So any observation of the leptonic CP violation, or demonstrating that CP is not a good symmetry of the leptons, can strengthen our belief in leptogenesis. In Ref.\cite{Ahn:2010cc} one 5-dimensional effective operator with respect to $\Lambda$ under $SU(2)\times U(1)\times A_4\times Z_2\times Z_4$ is introduced to have the aforementioned leptonic CP violation, and $A_{4}\times Z_{4}$ symmetries are broken after the assuming scalars develop VEVs with {\it ad hoc} constraints in the potential, interestingly, which opens the possibility to study an attractive mechanism of leptogenesis and LFV as well as to connect these with low-energy observables without contradicting $1\sigma$ results \cite{bari}, through symmetry breaking of $A_{4}$ in a radiative seesaw mechanism.

In this paper, to investigate the relation between neutrino parameters and high energy phenomenologies in the light of the deviation of TBM and the recent precision oscillation data, we focus on the normal hierarchical mass spectrum of light neutrino giving a successful leptogenesis at electroweak (EW) scale or more in a radiative seesaw mechanism. 
We find that, in the TBM limit, the branching ratio of $\mu(\tau)\rightarrow e\gamma$ and the $\mu-e$ conversion ratio $R_{\mu e}$ are going to be zero since these processes are sensitive to a deviation parameter $x$ which is proportional to the unknown mixing angle $\theta_{13}$, and interestingly that ${\rm Br}(\tau\rightarrow\mu\gamma)$ is mostly determined by the mass of DM $\bar{m}_{\eta}$ and a missing energy $m_-$ under $x\ll1$. Since the ratio of the branching ratios for ${\rm Br}(\mu\rightarrow e\gamma)/{\rm Br}(\tau\rightarrow\mu\gamma)$ is strongly dependent on the value of $\theta_{13}$, if future experiments of neutrino and LFV would measure the nonvanishing $U_{e3}$ and ${\rm Br}(\mu\rightarrow e\gamma)$ respectively,  ${\rm Br}(\tau\rightarrow \mu\gamma)$ can be predicted and at the same time the mass of DM ($\simeq$ leptogenesis scale) and a missing energy can be strongly bounded by this prediction. In addition, we show the magnitude of $R_{\mu e}$ is suppressed by $2-3$ orders compared to that of ${\rm Br}(\mu\rightarrow e\gamma)$.
Particularly, we show that, for example, with a missing energy $m_{-}=100$ eV the requirement for successful leptogenesis consistent with $\sin\theta_{13}\simeq{\cal O}(0.1)$ can be compatible with the existing constraint on ${\rm Br}(\mu\rightarrow e\gamma)$ if the mass of DM $\bar{m}_{\eta}\gtrsim285$ GeV.

The paper is organized as follows. In the next section, we briefly discuss the neutrino masses and mixings generated in a radiative seesaw mechanism and explain how the parameters are constrained by the low energy neutrino oscillation data. In Sec.~$\textrm{III}$, relation between neutrino parameters and LFV prediction with a successful leptogenesis is investigated. Then we give the conclusion in Sec.~$\textrm{IV}$.

\section{Low energy observables}

Unless flavor symmetries are assumed, particle masses and mixings are generally undetermined in gauge theory. To understand the present neutrino oscillation data we consider $A_{4}$ flavor symmetry for leptons, and simultaneously for the existence of DM, LHC signal and the BAU to be explained around EW scale or more we also introduce an extra discrete symmetry $Z_{2}$ in a radiative seesaw~\cite{Ma:2006km}, which could enhance LFV as reachable in near future experiments~\cite{Ma:2001mr}.
Especially, in Ref.\cite{Ahn:2010cc} a 5-dimensional operator is introduced in the lagrangian, which is invariant under $A_4\times Z_2\times Z_{4}$ to have non-zero low energy CP violation in neutrino oscillation and non-zero high energy cosmological CP violation which is responsible for BAU.
The technical details of the group are shown in Ref.\cite{Ahn:2010cc}.

The Yukawa interactions written in~\cite{Ahn:2010cc} can be replaced, after re-basing both charged lepton and heavy Majorana neutrino mass matrices to be diagonalized, by
 \begin{eqnarray}
 {\cal L}_{\rm Yuk} &=& \tilde{Y}_{\nu} \bar{\ell}_{L}\eta N_{i}+M^{d}_{R}\bar{N}_{i}(N_{i})^{c}+h.c
 \label{lagrangianA}
 \end{eqnarray}
where
 \begin{eqnarray}
  M^{d}_{R} =M{\rm Diag.}(a, 1, b)~,
  \label{MR3}
 \end{eqnarray}
where $a=\sqrt{1+\kappa^{2}+2\kappa\cos\xi}, b=\sqrt{1+\kappa^{2}-2\kappa\cos\xi}$, with real and positive mass eigenvalues,
and the couplings of $N_{i}$ with leptons and scalar $\eta$ is given as
 \begin{eqnarray}
 \tilde{Y}_{\nu}=
 g_{\nu}{\left(\begin{array}{ccc}
 -\frac{2+e^{i\phi}x}{\sqrt{6}} & \frac{1+e^{i\phi}x}{\sqrt{3}} &  \frac{e^{i\phi}x}{\sqrt{6}} \\
 \frac{1-e^{i\phi}x}{\sqrt{6}} &  \frac{1+\omega e^{i\phi}x}{\sqrt{3}}  &  \frac{i\sqrt{3}+e^{i\phi}x}{\sqrt{6}} \\
 \frac{1-e^{i\phi}x}{\sqrt{6}} &  \frac{1+\omega^{2} e^{i\phi}x}{\sqrt{3}}  &  \frac{-i\sqrt{3}+e^{i\phi}x}{\sqrt{6}}
 \end{array}\right)}Q_{\nu}~,
\label{Knu}
 \end{eqnarray}
where the diagonal matrix of heavy Majorana neutrino phase $Q_{\nu}={\rm Diag.}(e^{i\frac{\varphi_1}{2}},1,e^{i\frac{\varphi_2}{2}})$ with the phases
 \begin{eqnarray}
  \varphi_1=\tan^{-1}\Big(\frac{\kappa\sin\xi}{1+\kappa\cos\xi}\Big)~,~~~~
  \varphi_2=\tan^{-1}\Big(\frac{\kappa\sin\xi}{\kappa\cos\xi-1}\Big)~.
 \end{eqnarray}
Concerned with CP violation, we notice that the CP phases $\varphi_1,\varphi_2$ coming from $M_{R}$ as well as the CP phase $\phi$ from $Y_{\nu}$ obviously take part in low-energy CP violation, as you can see in Eq.~(\ref{meff3}).
\begin{figure*}[t]
\begin{tabular}{cc}
\includegraphics[width=6.0cm]{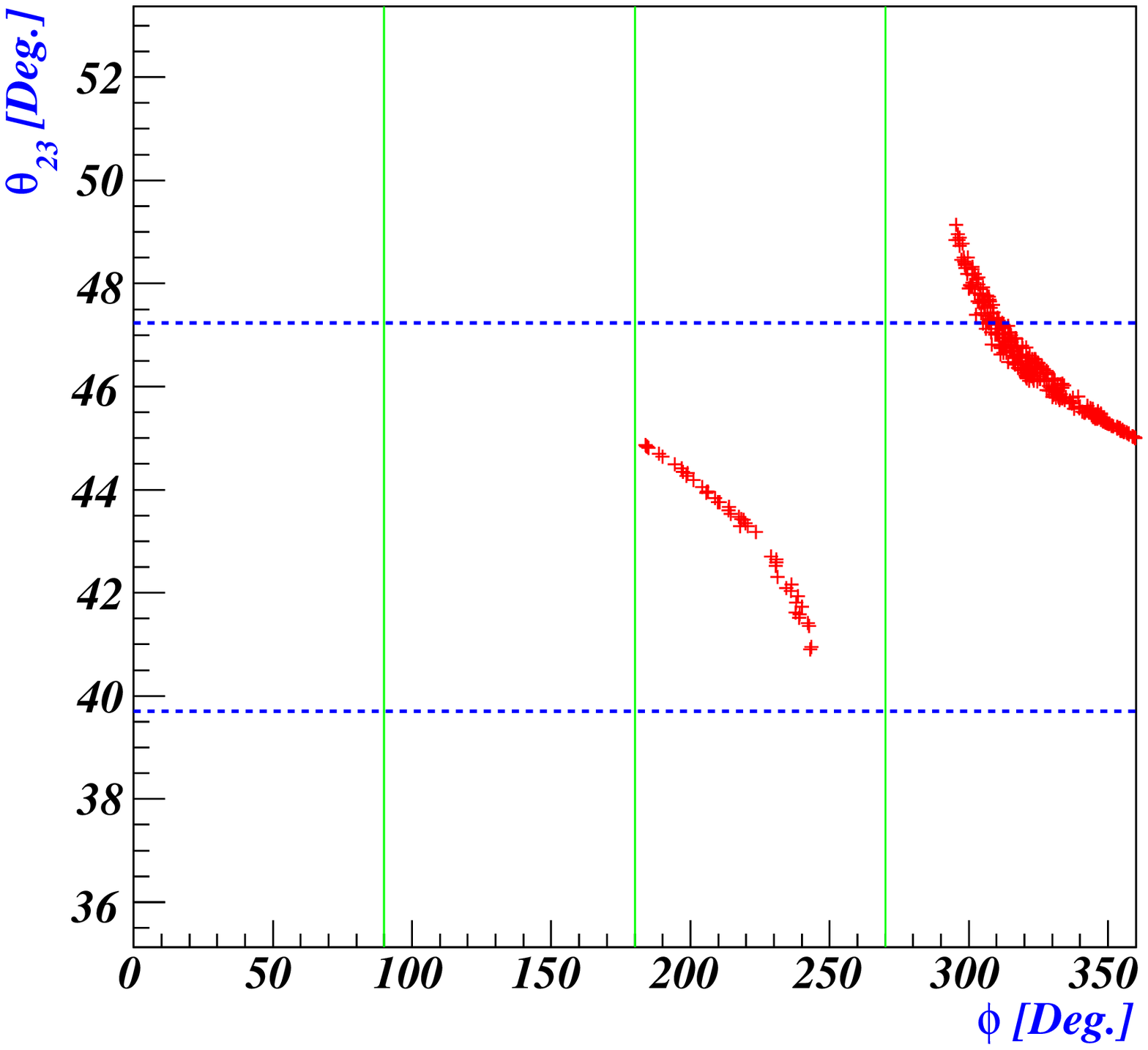}&
\includegraphics[width=6.0cm]{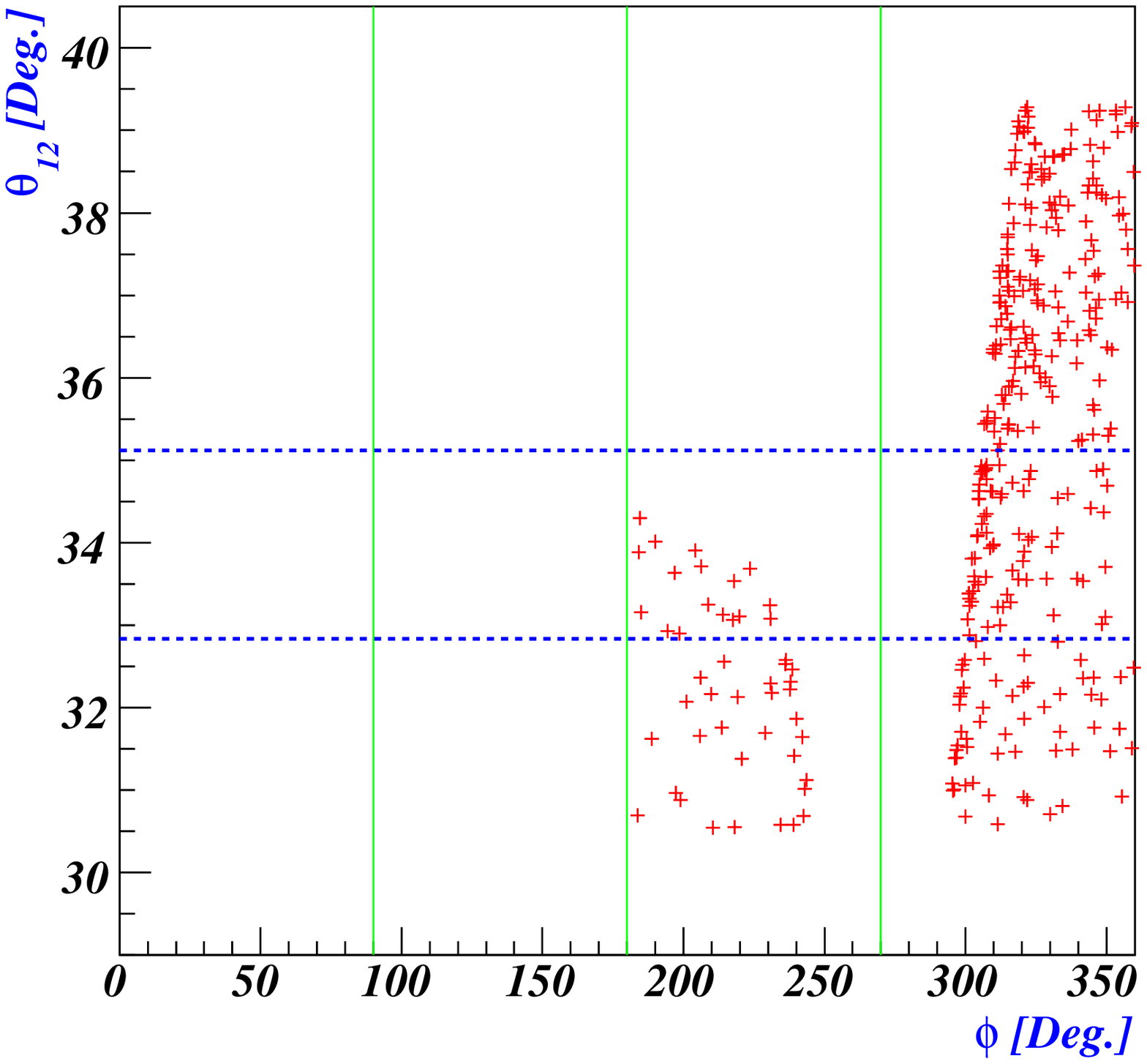}
\end{tabular}
\caption{\label{Fig1} Left-figure represents the atmospheric mixing angle $\theta_{23}$ over the phase $\phi$. Right-figure shows the solar mixing angle $\theta_{12}$ as a function of the parameter $\phi$. Here the horizontal dotted lines represent the experimental lower and upper bounds in $1\sigma$ of the mixing angle $\theta_{23}$.}
\end{figure*}
We assumed $M^{2}_{i}\gtrsim \bar{m}^{2}_{\eta}$, so the lightest $Z_{2}$-odd neutral particle of $\eta$ is stable and can be a candidate of DM. Therefore, after radiative-seesawing , for $M^{2}_{i}\gtrsim \bar{m}^{2}_{\eta}$ the effective light
neutrino mass matrix is obtained as
 \begin{eqnarray}
  m_{\nu} = \frac{\Delta m^{2}_{\eta}}{16\pi^{2}}\tilde{Y}_{\nu}M^{-1}_{R}\tilde{Y}^{T}_{\nu}~,
 \label{radseesaw3}
 \end{eqnarray}
where $\Delta m^{2}_{\eta}\equiv|m^{2}_{R}-m^{2}_{I}|=\mathcal{O}(\lambda_{\Phi\eta})\upsilon^{2}$ if $m_{R}(m_{I}) $ is the mass of $\eta^{0}_{R}(\eta^{0}_{I})$ and $m^{2}_{R(I)}=\bar{m}^{2}_{\eta}\pm\Delta m^{2}_{\eta}/2$\footnote{Actually, from potential lagrangian we can fully express the scalar masses $m_{\eta^{\pm}}, m_{R}, m_{I}$ and $\bar{m}_{\eta}$. However, for simplicity, these are expressed in terms of a relevant potential term.}, and $M_{R}={\rm Diag}(M_{r1},M_{r2},M_{r3})$ and $M_{ri}$ can be simplified as,
 \begin{eqnarray}
 M_{ri}\simeq\left\{\begin{array}{ll}
    2M_{i}, & \hbox{for $z_{i}\rightarrow1$} \\
    M_{i}\big[\ln z_{i}-1\big]^{-1} , & \hbox{for $z_{i}\gg1$.}
  \end{array}\right.
 \end{eqnarray}
 with $z_{i}=M^{2}_{i}/\bar{m}^{2}_{\eta}$.
The light neutrino mass matrix Eq.~(\ref{radseesaw3}) can not be diagonalized by the TBM mixing matrix
\begin{widetext}
 \begin{eqnarray}
  m_{\rm eff} =
  m_{0}U_{\rm TB}P_{\nu}{\left(\begin{array}{ccc}
  \frac{\omega_{1}}{a}+\frac{e^{2i\phi}x^{2}\omega_{2}}{2} & -\frac{xe^{i\phi}}{\sqrt{2}}(\frac{\omega_{1}}{a}+\omega_{2}) &  -\frac{e^{2i\phi}x^{2}\omega_{2}}{2} \\
  -\frac{xe^{i\phi}}{\sqrt{2}}(\frac{\omega_{1}}{a}+\omega_{2}) & \omega_{2}+\frac{x^{2}e^{2i\phi}}{2}(\frac{\omega_{1}}{a}+\frac{\omega_{3}}{b}) &  \frac{xe^{i\phi}}{\sqrt{2}}(\frac{\omega_{3}}{b}+\omega_{2}) \\
  -\frac{e^{2i\phi}x^{2}\omega_{2}}{2} & \frac{xe^{i\phi}}{\sqrt{2}}(\frac{\omega_{3}}{b}+\omega_{2}) &  \frac{\omega_{3}}{b}+\frac{e^{2i\phi}x^{2}\omega_{2}}{2}
 \end{array}\right)}P^{T}_{\nu}U^{T}_{\rm TB}~,
  \label{meff3}
 \end{eqnarray}
\end{widetext}
where $\omega_{i}=\frac{z_{i}}{1-z_{i}}[1+\frac{z_{i}\ln z_{i}}{1-z_{i}}]$, $m_{0}=\Delta m^{2}_{\eta}g^{2}_{\nu}/M16\pi^{2}$ and the TBM $U_{\rm TB}$ and the diagonal matrix of Majorana phase $P_{\nu}$ are
 \begin{eqnarray}
 U_{\rm TB} = {\left(\begin{array}{ccc}
  -\sqrt{\frac{2}{3}} &  \sqrt{\frac{1}{3}} &  0 \\
  \sqrt{\frac{1}{6}} &  \sqrt{\frac{1}{3}} &  \frac{1}{\sqrt{2}} \\
  \sqrt{\frac{1}{6}} &  \sqrt{\frac{1}{3}} &  \frac{-1}{\sqrt{2}}
 \end{array}\right)}~,~~~
 P_{\nu} = {\rm Diag.}(e^{i\frac{\varphi_1}{2}},1,e^{i\frac{\varphi_2+\pi}{2}})~.
 \label{TB}
 \end{eqnarray}
\begin{figure*}[b]
\begin{tabular}{cc}
\includegraphics[width=6.0cm]{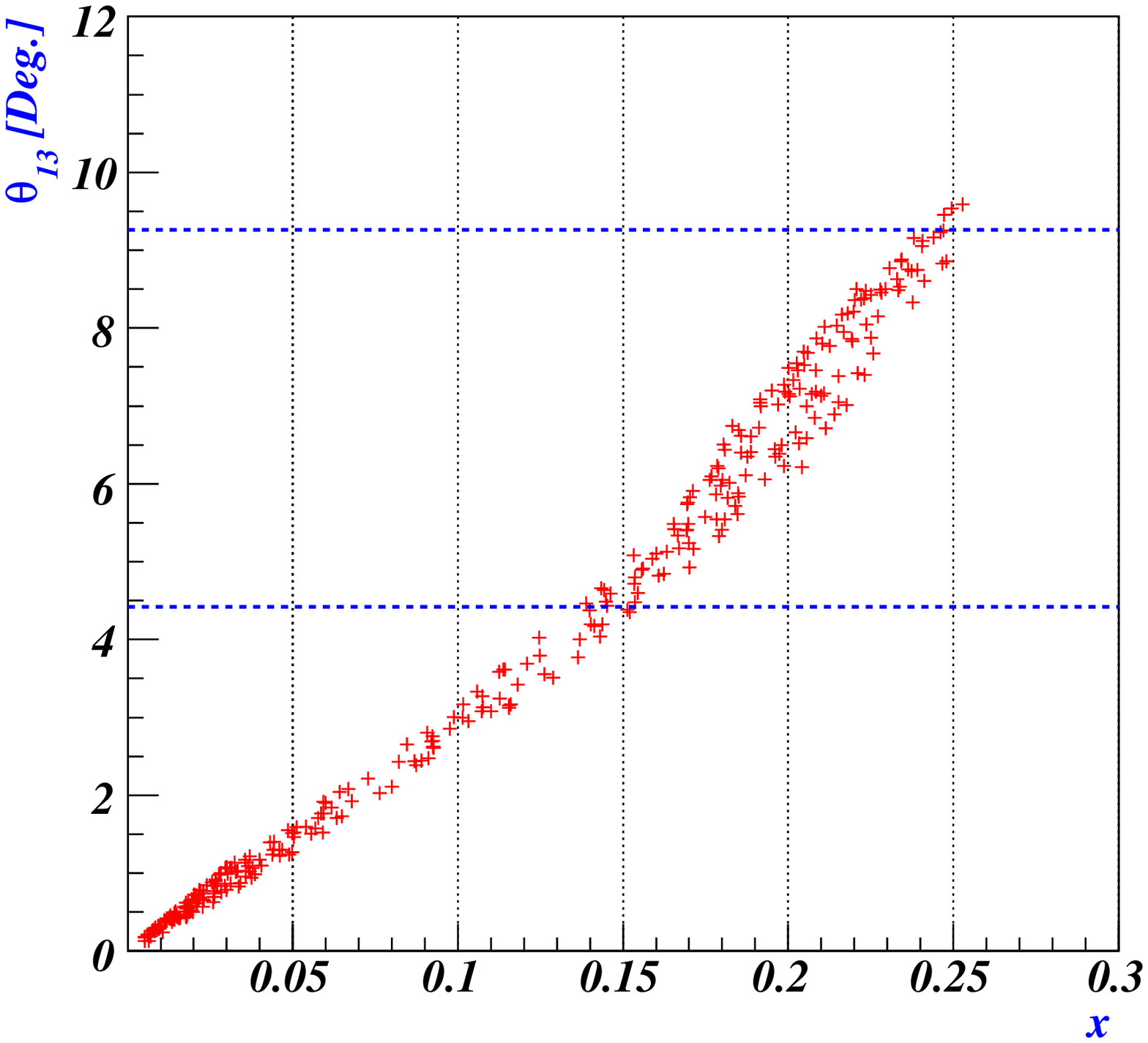}&
\includegraphics[width=6.0cm]{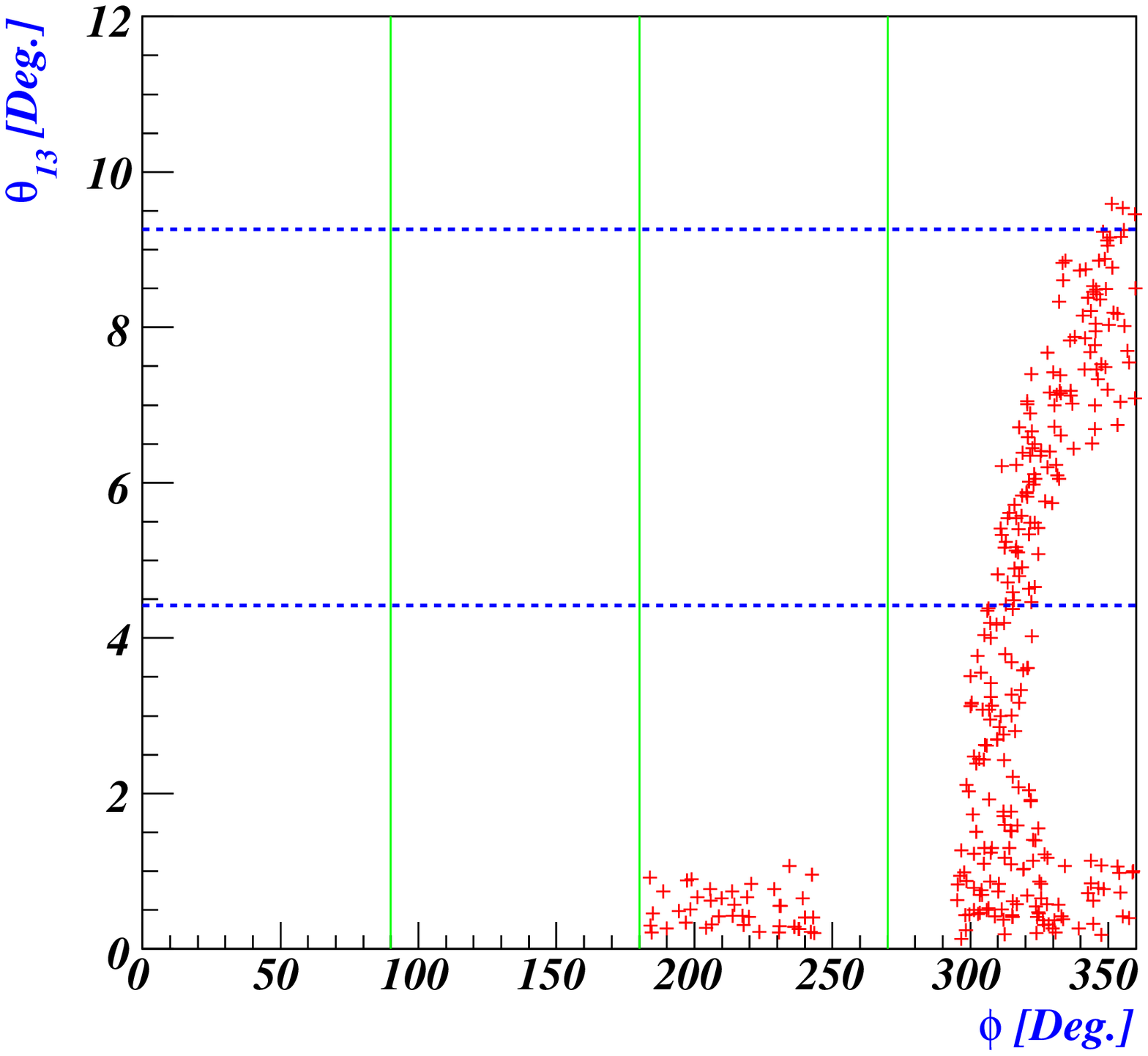}
\end{tabular}
\caption{\label{Fig2} Both figures represent the reactor angle $\theta_{13}$ over the dependence of the parameter $\phi$ and $x$, respectively. In the right figure, the horizontal dotted lines represent the experimental lower and upper bounds in $1\sigma$ of the mixing angle $\theta_{13}$.}
\end{figure*}
The neutrino mass matrix Eq.~(\ref{meff3}) represents that $\mu-\tau$ symmetry is broken by $x$, and can not be diagonalized by $U_{\rm TB}$ in Eq.~(\ref{TB}).
To diagonalize the above matrix Eq.~(\ref{meff3}), if we consider $m_{\rm eff}m^{\dag}_{\rm eff}$ one can obtain the masses and the mixing angles. For simplicity, we consider the case of $\varphi_{1,2}=0$ without a loss of generality.
Then, the light neutrino masses are given, up to first order of $x$, as
 \begin{eqnarray}
  |m_{1}|^{2} &\simeq& m^{2}_{0}\Big\{\frac{\omega^{2}_{1}}{a^{2}}-\frac{x}{4}\Big(\omega_{2}+\frac{\omega_{1}}{a}\Big)^{2}\cos\phi\Big\}~,\nonumber\\
  |m_{2}|^{2} &\simeq& m^{2}_{0}\Big\{\omega^{2}_{2}+\frac{x}{4}\Big(\omega_{2}+\frac{\omega_{1}}{a}\Big)^{2}\cos\phi\Big\}~, \nonumber\\
  |m_{3}|^{2} &\simeq& m^{2}_{0}\frac{\omega^{2}_{3}}{b^{2}}~.
  \label{eigenvalues}
 \end{eqnarray}
And, solar neutrino mixing is governed by
 \begin{widetext}
 \begin{eqnarray}
  \tan2\theta_{12} \simeq 2\sqrt{2}\frac{\omega^{2}_{2}-\frac{\omega^{2}_{1}}{a^{2}}+\frac{x}{2}(\frac{\omega_{1}}{a}+\omega_{2})^{2}\cos\phi+\frac{x^{2}}{2}\{(\frac{\omega_{3}}{b}+\omega_{2})^{2}+2\omega_{2}\frac{\omega_{3}}{b}\cos2\phi\}}{\omega^{2}_{2}
  -\frac{\omega^{2}_{1}}{a^{2}}-4x(\frac{\omega_{1}}{a}+\omega_{2})^{2}\cos\phi+\frac{x^{2}}{2}\{(\frac{\omega_{3}}{b}+\omega_{2})^{2}+2\omega_{2}\frac{\omega_{3}}{b}\cos2\phi\}}
  \label{theta12}
 \end{eqnarray}
 \end{widetext}
 which for $x=0$ agrees with the result of tri-bimaximal, i.e. $\tan2\theta_{12}=2\sqrt{2}$. Note here that in Eq.~(\ref{theta12}) the condition
$\omega^{2}_{2}-\frac{\omega^{2}_{1}}{a^{2}}+\frac{x^{2}}{2}\{(\frac{\omega_{3}}{b}+\omega_{2})^{2}+2\omega_{2}\frac{\omega_{3}}{b}\cos2\phi\}\gg|x(\frac{\omega_{1}}{a}+\omega_{2})^{2}\cos\phi|$
should be satisfied, in order for $\theta_{12}$ to be lie in the experimental bounds in Table-\ref{tab:data}. Especially, the right figure in Fig.~\ref{Fig1} shows the solar mixing angle $\theta_{12}$ as a function of $\phi$.
And the deviation from maximality of atmospheric neutrino mixing angle, $\Delta_{23}\equiv\theta_{23}+\pi/4$, comes out as
 \begin{eqnarray}
  \Delta_{23} \simeq \frac{6(\omega^{2}_{2}-\frac{\omega^{2}_{1}}{a^{2}})\sin\phi}{3(\omega^{2}_{2}-\frac{\omega^{2}_{1}}{a^{2}})\sin\phi+\sqrt{3}(\omega_{2}+\frac{\omega_{3}}{b})^{2}\cos\phi}~.
  \label{theta23}
 \end{eqnarray}
From Eq.~(\ref{theta23}), for the parameter $\kappa$ given by heavy neutrino mass ordering, $\Delta_{23}$ can be determined only by the parameter $\phi$, in which the values of $\phi$ at $\pi/2, 3\pi/2$ are not allowed by the experimental bounds of $\theta_{23}$, see also Fig.~\ref{Fig1}.
The unknown mixing angle $\theta_{13}$ and Dirac phase $\delta_{\rm CP}$ of $U_{\rm PMNS}$ can be obtained approximately, for $x\ll1$, as
 \begin{eqnarray}
  \theta_{13}&\simeq& -x\sqrt{\frac{3}{2}}\frac{(\frac{\omega^{2}_{3}}{b^{2}}-\omega^{2}_{2})\sin\phi\cos\delta_{CP}+(\omega_{2}+\frac{\omega_{3}}{b})^{2}\cos\phi\sin\delta_{CP}}{3\frac{\omega^{2}_{3}}{b^{2}}-\omega^{2}_{2}-\frac{\omega^{2}_{1}}{a^{2}}}~,\nonumber\\
  \delta_{CP} &\simeq& \tan^{-1}\Big(\frac{\omega_{3}+b\omega_{2}}{\omega_{3}-b\omega_{2}}\cot\phi\Big)~,
  \label{theta13}
 \end{eqnarray}
in which $\theta_{13}$ is closely proportional to the size of $x$ and also related with $\phi$, as can be seen in Fig.~\ref{Fig2}, and $\delta_{CP}$ is mainly determined by the phase $\phi$, when the parameters $\omega_{i}, b$ are fixed by the mass ordering of heavy Majorana neutrinos according to the light neutrino mass spectrum.
From Eq.~(\ref{theta23}) and Eq.~(\ref{theta13}), we see that the deviation of $\theta_{23}$ is linked to $\theta_{13}$ through phase $\phi$, not through the parameter $x$.  Interesting points are that the deviation of $\theta_{12}$ from tri-bimaximal is closely related with $\theta_{13}$ through the parameters $x$ and $\phi$, and the deviation of $\theta_{23}$ from maximality is governed by the phase $\phi$ which is related with $\delta_{CP}$ in Eq.~(\ref{theta13}), if the parameter $\kappa$ is determined by heavy neutrino mass ordering.

Because of the observed hierarchy $|\Delta m^{2}_{32}|\gg\Delta m^{2}_{21}$, and the requirement of MSW resonance for solar neutrinos, there are two possible neutrino mass spectrum: (i) $m_{1}<m_{2}<m_{3}$ (normal mass spectrum) which corresponds to $\frac{\omega_{1}}{a}<\omega_{2}<\frac{\omega_{3}}{b}$ and (ii) $m_{3}<m_{1}<m_{2}$ (inverted mass spectrum) which corresponds to $\frac{\omega_{3}}{b}<\frac{\omega_{1}}{a}<\omega_{2}$. The solar and atmospheric mass-squared differences defined as $\Delta m^{2}_{ij}\equiv m^{2}_{i}-m^{2}_{j}$ are given by
 \begin{eqnarray}
  \Delta m^{2}_{21} \simeq m^{2}_{0}\Big\{\omega^{2}_{2}-\frac{\omega^{2}_{1}}{a^{2}}+\frac{x}{2}\big(\omega_{2}+\frac{\omega_{1}}{a}\big)^{2}\cos\phi\Big\}~,
  \label{massSqd1}
 \end{eqnarray}
 \begin{eqnarray}
  \Delta m^{2}_{32}\simeq m^{2}_{0}\Big\{\frac{\omega^{2}_{3}}{b^{2}}-\omega^{2}_{2}-\frac{x}{4}\big(\omega_{2}+\frac{\omega_{1}}{a}\big)^{2}\cos\phi\Big\}~,
  \label{massSqd2}
 \end{eqnarray}
in which, from the neutrino oscillation experiments we know that $\Delta m^{2}_{21}$ is positive and dictates $\omega_{2}>\omega_{1}/a$ with the second term being sufficiently small $x$ in Eq.~(\ref{massSqd1}). In order for a leptogenesis to be successfully implemented at or around EW scale in our scenario (see, Ref.\cite{Ahn:2010cc}), we considered the case $M^{2}_{\rm lightest}\simeq \bar{m}^{2}_{\eta}$ where the $M_{\rm lightest}$ is the lightest of the heavy Majorana neutrino due to a strong wash-out. Moreover, since the hierarchical mass ordering of the heavy Majorana neutrino masses could give a successful leptogenesis, we consider here the case  $M_{1,2}\gg M_{3}$ ($a>1\gg b$ with $\xi=0$): this case corresponds to the normal hierarchical mass spectrum with $b\rightarrow0$ i.e. $\kappa\simeq1$. Using $\omega_{1}\simeq2\ln\frac{2}{b}-1$, $\omega_{2}\simeq-2\ln b-1$ and $\omega_{3}\simeq\frac{1}{2}$, the ratio of the mass squared differences defined by $R\equiv\Delta m^{2}_{21}/|\Delta m^{2}_{32}|$ which is around $R\simeq3\times10^{-2}$ for the best-fit values of the solar and atmospheric mass squared differences, which is given by
 \begin{eqnarray}
  R\approx b^{2}(4\omega^{2}_{2}-\omega^{2}_{1})~,
  \label{RM3}
 \end{eqnarray}
 where the equality roughly can be given under $1\gg x,b$. Note here that using the best-fit value of $R(\simeq3\times10^{-2})$ and Eq.~(\ref{RM3}) one can roughly determine the size of the parameter $b$, i.e $b\simeq0.01$.
Since $m_{0}=\Delta m^{2}_{\eta}g^{2}_{\nu}/16\pi^{2}M$ as defined in Eq.~(\ref{meff3}), the value of $g_{\nu}$ depends on the magnitude of $m_{-}\equiv|m_{R}-m_{I}|$ in the case that $m_{0}$ is determined as
 \begin{eqnarray}
  m_{0}\simeq\frac{m_-bg^2_{\nu}}{8\pi^2}\simeq m_{3}\frac{b}{\omega_{3}}
 \label{mo1}
 \end{eqnarray}
where $\Delta m^{2}_{\eta}=m_{+}m_{-}$ and $m_+\equiv|m_{R}+m_{I}|\simeq2\bar{m}_{\eta}$ are used.
Since all new scalars $\eta^{\pm}, \eta^{0}_{R}, \eta^{0}_{I}$ carry a $Z_{2}$ odd quantum number and only couple to Higgs boson and electroweak gauge bosons of the SM, they can be produced in pairs through the SM gauge bosons $W^{\pm}, Z$ or $\gamma$.  Once produced, $\eta^{\pm}$ will decay into $\eta^{0}_{R,I}$ and a virtual $W^{\pm}$, then $\eta^0_{I}$ subsequently becomes $\eta^0_{R}$ + $Z$-boson, which will decay a quark-antiquark or lepton-antilepton pair. Here, for example, the mass hierarchy $m_{\eta^{\pm}}>m_{I}>m_{R}$ is assumed. That is, the stable $\eta^{0}_{R}$ appears as missing energy in the decays of $\eta^{\pm}\rightarrow\eta^{0}_{I}l^{\pm}\nu$ with the subsequent decay $\eta^{0}_{I}\rightarrow\eta^{0}_{R}l^{\pm}l^{\mp}$, which can be compared to the direct decay $\eta^{\pm}\rightarrow\eta^{0}_{R}l^{\pm}\nu$ to extract the masses of the respective particles. Therefore, if the signal of $m_{-}$ and $m_{+}$ in LHC are measured, i.e. $\bar{m}_{\eta}\simeq M_{\rm lightest}\gtrsim {\rm electroweak ~scale}$, the lightest of heavy Majorana neutrinos can be decided. As will be shown later, these $m_{\pm}$ are strongly dependent on the LFV of $\tau\rightarrow\mu\gamma$, which means finding $m_{\pm}$ is the search of the branching ratio of $\tau\rightarrow\mu\gamma$ and vice versa searching the branching ratio of $\tau\rightarrow\mu\gamma$ can strongly constrain the values of $m_{\pm}$.

{\it Numerical Analysis:}
as can be seen in Eqs.~(\ref{eigenvalues}-\ref{theta13}), three neutrino masses, three mixing angles and a CP phase are presented in terms of five independent parameters $m_{0},\kappa$(or $a,b$), $x, \phi$. Interestingly, if we focus on the case $M_{1,2}\gg M_{3}$ which gives a normal hierarchical mass spectrum of light neutrino, the values of parameters $m_{0},\kappa$(or $a,b$) can be determined as $m_{0}\simeq0.001,\kappa\simeq1$, independent of the mass scale of heavy Majorana neutrinos, which in turn indicates that three neutrino masses, three mixing angles and a CP phase only can be corrected by the parameters $x$ and $\phi$. However, as will be shown later, since both LFV in Eq.~(\ref{LFV}) and a leptogenesis for a fixed value $\delta_{N_3}$ in Eq.~(\ref{degeneracy}) are dependent on the mass of the lightest heavy neutrino $N_{3}$ and the missing energy $m_{-}$ as Eq.~(\ref{mo1}), for convenience, we first fix the value of heavy Majorana neutrino with second one being to be $M_{2}\equiv M=17$ TeV and $m_{-}=100$ eV. Then, we impose the current experimental results on neutrino masses and mixings into the hermitian $m^{\dag}_{\rm eff}m_{\rm eff}$ and varying all the parameter spaces $\{\kappa, \phi, x\}$:
 \begin{eqnarray}
  0.98\lesssim\kappa\lesssim1.02~,~~~~~0\leq\phi\leq2\pi~,~~~~~0.005\leq x<0.4~.
 \end{eqnarray}
As a result of the numerical analysis, Fig.~\ref{Fig1} and Fig.~\ref{Fig2} show how the mixing angles $\theta_{12},\theta_{13}$ and $\theta_{23}$ depend on the parameter $\phi$, which are explained in the approximate analysis Eqs.(\ref{theta12}-\ref{theta13}). The left figure in Fig.~\ref{Fig1} shows the dependence of phase $\phi$ on the atmospheric mixing angle, in which $\theta_{23}$ favors $\theta_{23}<\pi/4$ and $\theta_{23}>\pi/4$ in the region $\pi<\phi<3\pi/2$ and $3\pi/2<\phi<2\pi$, respectively, and the right figure in Fig.~\ref{Fig1} represents the solar mixing angle $\theta_{12}$ as a function of $\phi$; in the region $\pi<\phi<3\pi/2$, $\theta_{12}$ favors $\theta_{12}<35.3^{\circ}$. Fig.~\ref{Fig2} represents that how the reactor angle $\theta_{13}$ depends on the parameters $x$ and $\phi$, as can be seen in Eq.~(\ref{theta13}), where $\theta_{13}$ prefers to very small values less than $1^{\circ}$ in the region $\pi<\phi<3\pi/2$, on the other hand, $0\lesssim\theta_{13}\lesssim10^{\circ}$ for $3\pi/2<\phi<2\pi$. Finally, the Dirac CP phase can be determined by Eq.~(\ref{theta13}) with the parameter $\phi$. In addition, to show flavor effects in leptogenesis as well as the dependence of the mass of DM, we use $M_{2}\equiv M=16, 31,82$TeV and $m_{-}=100$ eV as inputs, which will be shown in Fig.~\ref{Fig6}.

\section{Lepton Flavor Violation and Leptogenesis}
\begin{figure*}[t]
\begin{tabular}{c}
\includegraphics[width=6.0cm]{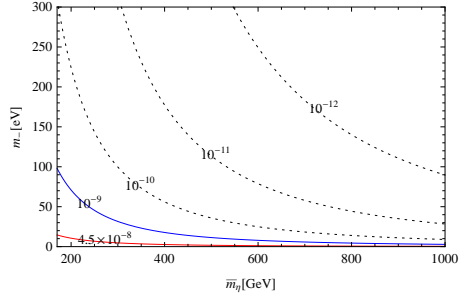}
\end{tabular}
\caption{\label{Fig4} Contour plot of ${\rm Br}(\tau\rightarrow \mu\gamma)$ as the functions of $m_{-}$ and $\bar{m}_{\eta}$, where the values in the plot denote the branching ratio of ${\rm Br}(\tau\rightarrow \mu\gamma)$. Here the phase $\phi$ does not affect considerably, and $\phi=340^{\circ}$ is taken.}
\end{figure*}

\begin{figure}[t]
\begin{tabular}{cc}
\includegraphics[width=6.0cm]{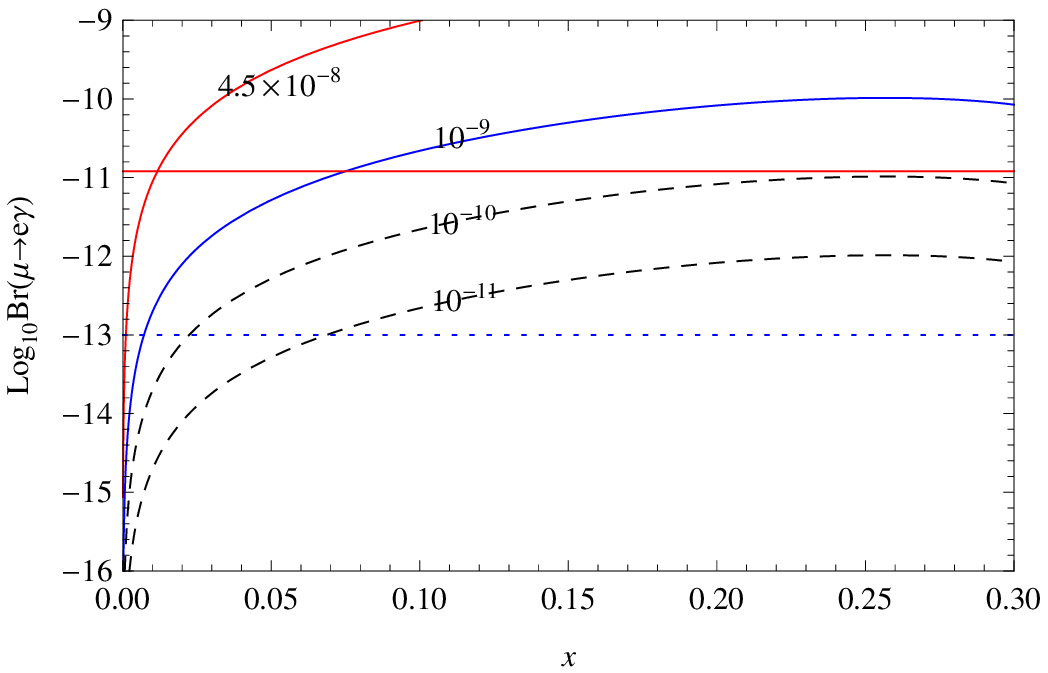}&
\includegraphics[width=6.0cm]{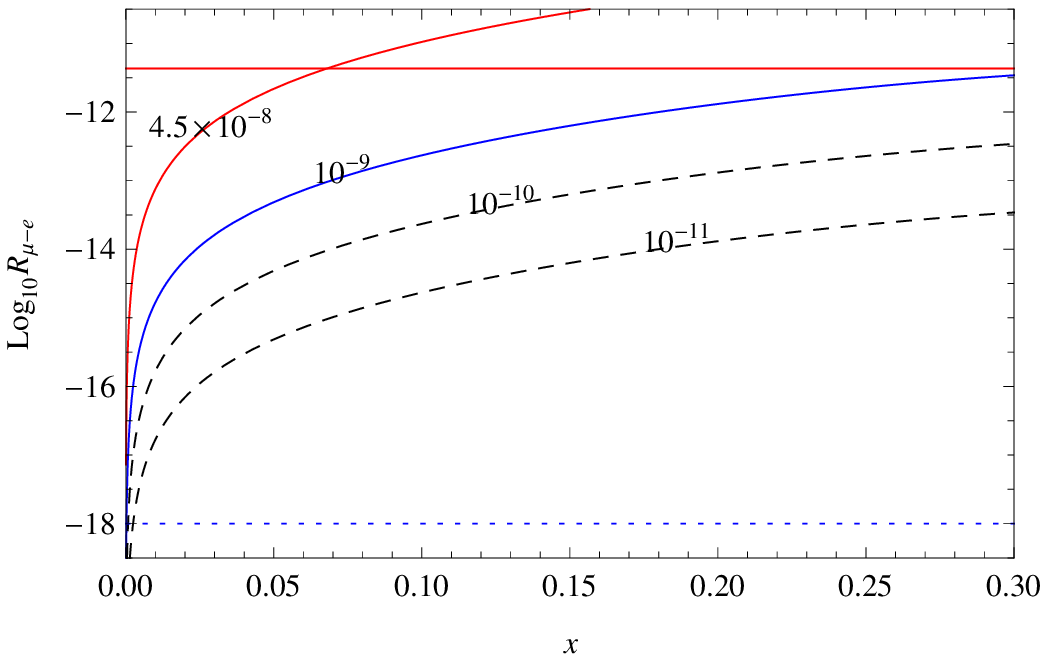}
\end{tabular}
\caption{\label{Fig5}  (Left-plot) Predictions of ${\rm Br}(\mu\rightarrow e\gamma)$ as the functions of $x$ and ${\rm Br}(\tau\rightarrow \mu\gamma)$, where the current (expected) bounds for $\mu\rightarrow e\gamma$ are shown by the solid (dotted) horizontal lines. (Right-plot) Predictions of $R_{\mu e}$ as the functions of $x$ and ${\rm Br}(\tau\rightarrow \mu\gamma)$, where the current (expected) bounds for $R_{\mu e}$ are shown by the solid (dotted) horizontal line. Here the values in both plots denote the branching ratio of ${\rm Br}(\tau\rightarrow \mu\gamma)$, and the phase $\phi$ does not affect considerably and here $\phi=340^{\circ}$ is taken.}
\end{figure}

The existence of the flavor neutrino mixing implies that the individual lepton charges, $L_{\alpha}, \alpha=e,\mu,\tau$ are not conserved~\cite{Bilenky:1987ty} and processes like $\ell_{\alpha}\rightarrow\ell_{\beta}\gamma$ should take place. Experimental discovery of lepton rare decay processes $\ell_{\alpha}\rightarrow\ell_{\beta}\gamma$ is one of smoking gun signals of physics beyond the SM; thus several experiments have been developed to detect LFV processes. The present experimental upper bounds are given at $90\%$ C.L.~\cite{Brooks:1999pu} as
 \begin{eqnarray}
  {\rm Br}(\mu\rightarrow e\gamma) &\leq&1.2\times10^{-11}~,~~~~~~{\rm Br}(\tau\rightarrow \mu\gamma)\leq4.5\times10^{-8}~,\nonumber\\
  {\rm Br}(\tau\rightarrow e\gamma) &\leq&1.2\times10^{-7}~.
 \label{exp}
 \end{eqnarray}
One-loop diagrams to the one for neutrino masses contribute to the lepton flavor violating processes like $\ell_{\alpha}\rightarrow\ell_{\beta}\gamma$~$(\alpha,\beta=e,\mu,\tau)$, whose branching ratio is estimated as~\cite{Ma:2001mr}
 \begin{eqnarray}
  {\rm Br}(\ell_{\alpha}\rightarrow\ell_{\beta}\gamma)=\frac{3\alpha_{e}}{64\pi(G_{F}\bar{m}^{2}_{\eta})^{2}}|B_{\alpha\beta}|^{2}{\rm Br}(\ell_{\alpha}\rightarrow\ell_{\beta}\bar{\nu}_{\beta}\nu_{\alpha})
 \label{LFV}
 \end{eqnarray}
where $\alpha_{e}\simeq1/137$ and $G_{F}$ is the Fermi constant, and $B_{\alpha\beta}$ is given by
 \begin{eqnarray}
  B_{\alpha\beta}=\sum^{3}_{i=1}
  \tilde{Y}_{\alpha i}\tilde{Y}^{\ast}_{\beta i}F_{2}(z_{i})~,
 \end{eqnarray}
in which $F_{2}(z_{i})$ is given by
 \begin{eqnarray}
  F_{2}(z_{i})=\frac{1-6z_{i}+3z^{2}_{i}+2z^{3}_{i}-6z^{2}_{i}{\rm ln}z_{i}}{6(1-z_{i})^{4}}
 \end{eqnarray}
 with $F_{2}(1)=1/12$. Taking the case $M_{1}\simeq2M,~M_{2}\simeq M$ and $M_{3}\simeq0.01M\gtrsim\bar{m}_{\eta}$ into account, the function $F_{2}(z_{i})$ can have the values $1/12$ $(z_{3}=1)$, $3.3\times10^{-5}$ $(z_{2}=10^{4})$ and $8.3\times10^{-6}$ $(z_{1}=4\times10^{4})$, for $N_{1}, N_{2}$ and $N_{3}$, respectively, which indicates only the lightest heavy Majorana neutrino among the heavy neutrinos can contribute the branching ratio of LFV. Then, the expressions of $|B_{\alpha\beta}|^{2}$ relevant for $\tau\rightarrow\mu\gamma$, $\tau\rightarrow e\gamma$ and $\mu\rightarrow e\gamma$, respectively, are approximately given as
\begin{widetext}
\begin{eqnarray}
  |B_{\tau\mu}|^{2}&\simeq& 201\Big(\frac{m_{3}}{m_{-}}\Big)^{2}\Big\{\frac{1}{4}-x+\frac{x^{2}}{6}(9+\cos2\phi)\Big\}~,\nonumber\\
  |B_{\tau e}|^{2} &\simeq& 5.6\times x^{2}\Big(\frac{m_{3}}{m_{-}}\Big)^{2}\Big\{3-2x(3+\sqrt{3}\sin\phi)+2x^{2}(2+\sqrt{3}\sin\phi)\Big\}~,\nonumber\\
  |B_{\mu e}|^{2}  &\simeq& 5.6\times x^{2}\Big(\frac{m_{3}}{m_{-}}\Big)^{2}\Big\{3-2x(3-\sqrt{3}\sin\phi)+2x^{2}(2-\sqrt{3}\sin\phi)\Big\}~,
 \label{Bab}
 \end{eqnarray}
\end{widetext}
where $g^{2}_{\nu}\simeq16\pi^{2}m_{3}/m_{-}$ in Eq.~(\ref{mo1}) is used. Note here that for $g_{\nu}\lesssim1$ the lower bound of $m_{-}$ is obtained as $8{\rm eV}\lesssim m_{-}$. And, the size of $m_{-}$ is also crucial for leptogenesis to be successfully implemented, which needs to be $m_{-}\simeq{\cal O}(100)$ eV, which will be shown later.
From Eq.~(\ref{LFV}) and Eq.~(\ref{Bab}) the LFV branching ratios can be simplified as
\begin{eqnarray}
  {\rm Br}(\ell_{\alpha}\rightarrow \ell_{\beta}\gamma)&\simeq&8.0\times10^{5}r_{\alpha\beta}\frac{|B_{\alpha\beta}|^{2}}{\bar{m}^{4}_{\eta}}{\rm GeV}^{4}~,
 \label{LFV1}
 \end{eqnarray}
where $r_{\mu e}=1.0, r_{\tau e}=0.1784$ and $r_{\tau\mu}=0.1736$, which indicates that the LFV branching ratios depends on $\bar{m}_{\eta}$ and on the neutrino parameters in a flavor dependent manner. In the TBM limit, ${\rm Br}(\mu\rightarrow e\gamma)$ and ${\rm Br}(\tau\rightarrow e\gamma)$ in Eqs.~(\ref{Bab},\ref{LFV1}) are going to be zero since these processes are sensitive to the parameter $x$ which represents the deviation from the TBM. Since the mixing element $|U_{e3}|$ is proportional to the deviation parameter $x$ which is constrained to be small but may still be nonzero, as can be seen in Eqs.~(\ref{Bab},\ref{LFV1}), predictions of the processes $\mu\rightarrow e\gamma, \tau\rightarrow e\gamma$ highly depend on the parameter $x(\ll1)$, however $\tau\rightarrow\mu\gamma$ does not have so large dependence on it. Therefore, ${\rm Br}(\tau\rightarrow\mu\gamma)$ is mostly determined by $\bar{m}_{\eta}$ and $m_-$, and due to ${\rm Br}(\tau\rightarrow\mu\gamma)\gg{\rm Br}(\tau\rightarrow e\gamma)$ we focus on $\tau\rightarrow\mu\gamma$ between the tau decay processes :
\begin{widetext}
\begin{eqnarray}
  {\rm Br}(\tau\rightarrow \mu\gamma)\simeq7\times10^{4}\Big(\frac{{\rm eV}}{m_{-}}\Big)^{2}\Big(\frac{{\rm GeV}}{\bar{m}_{\eta}}\Big)^{4}\Big\{\frac{1}{4}-x+\frac{x^{2}}{6}(9+\cos2\phi)\Big\}~,
 \label{taumu}
 \end{eqnarray}
\end{widetext}
which is constrained by the current upper bound Eq.~(\ref{exp}). Future $B$-factories would also greatly reduce the $\tau$ decay upper bounds~\cite{Hewett:2004tv}. In our analysis, we conservatively adopt
${\rm Br}(\tau\rightarrow \mu\gamma)\lesssim10^{-9}$ and ${\rm Br}(\tau\rightarrow e\gamma)\lesssim10^{-9}$ as upcoming upper bounds of LFV branching ratios. The contour for ${\rm Br}(\tau\rightarrow \mu\gamma)$ as the functions of $m_{-}$ and $\bar{m}_{\eta}$ is plotted in Fig.~\ref{Fig4}, where the values appeared in the figure represent the branching ratio of $\tau\rightarrow \mu\gamma$. Interestingly enough, Fig.~\ref{Fig4} represents if LFV $\tau\rightarrow \mu\gamma$ is measured in the future B-factories~\cite{Hewett:2004tv} the DM/LHC signal $\bar{m}_\eta$ can be bounded by $m_{-}$ in our scenario, where the influence of $|U_{e3}|$ is not considerable and $x=0.2,\phi=340^{\circ}$ is taken.
We see clearly that the branching ratio for $\tau\rightarrow \mu\gamma$ becomes higher for a fixed $m_{-}$ as $\bar{m}_{\eta}$ gets smaller.
As the parameter $x$ goes to zero, both $B_{\tau e}$ and $B_{\mu e}$ go to zero,  which means that the processes $\mu\rightarrow e\gamma, \tau\rightarrow e\gamma$ are strongly bounded by the unknown mixing angle $\theta_{13}$. Moreover, the contributions of the phase $\phi$ to $|B_{\tau e(\tau\mu,\mu e)}|^{2}$ are not considerable due to $x\ll1$ under the parameter spaces compatible with the current neutrino data Table-\ref{tab:data}. From Eqs.~(\ref{Bab},\ref{LFV1}) it is clear that the most of constraints comes from ${\rm Br}(\mu\rightarrow e\gamma)$ which is sensitive to the value of $|U_{e3}|$ as well as the mass of DM and the missing energy $m_{-}$.
Since the bound for ${\rm Br}(\mu\rightarrow e\gamma)$ is most severe, ${\rm Br}(\mu\rightarrow e\gamma)/{\rm Br}(\tau\rightarrow \mu(e)\gamma)$ must be sufficiently suppressed as $10^{-2}-10^{-5(6)}$ in order to observe both $\mu$ and $\tau$ decay processes.
The branching ratios for the three transitions $\mu\rightarrow e\gamma$, $\tau\rightarrow\mu\gamma$ and $\tau\rightarrow e\gamma$ are
${\rm Br}(\tau\rightarrow\mu\gamma)\gg{\rm Br}(\tau\rightarrow e\gamma)\simeq{\rm Br}(\mu\rightarrow e\gamma),$
in which the ratio of the branching ratios for ${\rm Br}(\mu\rightarrow e\gamma)$ and ${\rm Br}(\tau\rightarrow\mu\gamma)$ is only dependent on the parameters $x$ and $\phi$, which is approximately given as
\begin{widetext}
 \begin{eqnarray}
  \frac{{\rm Br}(\mu\rightarrow e\gamma)}{{\rm Br}(\tau\rightarrow\mu\gamma)}\simeq0.64x^{2}\frac{3-2x(3-\sqrt{3}\sin\phi)+2x^{2}(2-\sqrt{3}\sin\phi)}{1-4x+\frac{2x^{2}}{3}(9+\cos2\phi)}~.
 \label{RatioBmeBtm}
 \end{eqnarray}
\end{widetext}
Note here that this ratio is independent of the parameters $m_{-}$ and $\bar{m}_{\eta}$, and drastically changed by the parameter $x$ or $|U_{e3}|$. If the unknown mixing angle $|U_{e3}|$ is measured the ratio of Eq.~(\ref{RatioBmeBtm}) can be determined, which indicates if the branching ratio of $\mu\rightarrow e\gamma$ is measured the branching ratio $\tau\rightarrow \mu\gamma$ is predicted and then DM mass and missing energy can be determined. Given the present experimental bound ${\rm Br}(\mu\rightarrow e\gamma)<1.2\times10^{-11}$~\cite{Brooks:1999pu} and the experimental data of neutrino at $1\sigma$ in Table-\ref{tab:data}, Eq.~(\ref{RatioBmeBtm}) implies that $\tau\rightarrow\mu\gamma$ has rates much below the present and expected future sensitivity~\cite{Hewett:2004tv}.
The left plot of Fig.~\ref{Fig5} shows the predictions of ${\rm Br}(\mu\rightarrow e\gamma)$ as the functions of $x$ and ${\rm Br}(\tau\rightarrow\mu\gamma)$, where the values in the figure denote the branching ratio of ${\rm Br}(\tau\rightarrow \mu\gamma)$, the phase $\phi$ does not affect considerably and here $\phi=340^{\circ}$ is taken in Eq.~(\ref{RatioBmeBtm}).
If a relatively large value of reactor angle $|U_{e3}|=0.126$ (best-fit) is measured in near future, the left plot of Fig.~\ref{Fig5} shows the constraints from the upper bound of $\mu\rightarrow e\gamma$ requires the one of $\tau\rightarrow\mu\gamma$ should be less than $10^{-10}$. From Fig.~\ref{Fig5}, one can realize that future LFV searches with $|U_{e3}|$ measurements give us implications for the mass of DM. For example, if future experiments discover both ${\rm Br}(\mu\rightarrow e\gamma)$ and $|U_{e3}|$, then ${\rm Br}(\tau\rightarrow \mu\gamma)$ can be predicted, which in turn means $m_{-}$ and the DM mass are strongly constrained.  Note here that ${\rm Br}(\tau\rightarrow e\gamma)$ can be straightforwardly obtained from ${\rm Br}(\mu\rightarrow e\gamma)$, as the relationship ${\rm Br}(\tau\rightarrow e\gamma)={\rm Br}(\mu\rightarrow e\gamma)\cdot{\rm Br}(\tau\rightarrow e\bar{\nu}_{e}\nu_{\tau})$ with ${\rm Br}(\tau\rightarrow e\bar{\nu}_{e}\nu_{\tau})\approx17.84\%$~\cite{Amsler:2008zzb} hold.

It is worth while to mention about the $\mu-e$ conversion in two nuclei $^{27}_{13}$Al and $^{48}_{22}$Ti, proposed by Mu2e experiment at Fermilab~\cite{Mu2e} and PRIME experiment at J-PARK~\cite{PRIME}, aiming for sensitivities of $10^{-16}-10^{-18}$, respectively. And there is the  current bound $\Gamma(\mu{\rm Ti}\rightarrow e{\rm Ti})/\Gamma(\mu{\rm Ti}\rightarrow capture)<4.3\times10^{-12}$~\cite{Amsler:2008zzb}. The formula of the $\mu-e$ conversion ratio $R_{\mu e}$ given in Ref.~\cite{Ma:2001mr} can be simplified in our scenario as
 \begin{eqnarray}
 R_{\mu e}\simeq\frac{\alpha^{5}_{e}m^9_{\mu}Z^{4}_{\rm eff}Z|\bar{F}_{p}(p_{e})|^2}{2592\Gamma_{\rm capt}q^4\bar{m}^4_\eta}\Big(\frac{m_3}{m_-}\Big)^{2}x^{2}\{3+2\sqrt{3}x\sin\phi+x^2\}~,
 \label{Rmue}
 \end{eqnarray}
where $q^{2}\approx-m^{2}_{\mu}$, for $^{27}_{13}$Al, $Z_{\rm eff}=11.5, \bar{F}_{p}=0.64, \Gamma_{\rm capt}=4.64079\times10^{-19}$ GeV, and for $^{48}_{22}$Ti, $Z_{\rm eff}=17.6, \bar{F}_{p}=0.54, \Gamma_{\rm capt}=1.70422\times10^{-18}$ GeV. As can be seen in Eq.~(\ref{Rmue}), we can easily find that the ratio $R_{\mu e}$ strongly depends on the parameters $\bar{m}_{\eta},m_{-}$ and $x$(or $\theta_{13}$) in the same way with ${\rm Br}(\mu\rightarrow e\gamma)$. And comparing between $R_{\mu e}$ and ${\rm Br}(\tau\rightarrow\mu\gamma)$, the ratio of between them is simplified as
 \begin{eqnarray}
 \frac{R_{\mu e}}{{\rm Br}(\mu\rightarrow e\gamma)}\simeq4.7(8.4)\times10^{-3}\frac{3+2\sqrt{3}x\sin\phi+x^2}{3-2x(3-\sqrt{3}\sin\phi)+2x^{2}(2-\sqrt{3}\sin\phi)}~,
 \label{Rmue1}
 \end{eqnarray}
where the factors $4.7$ and $8.4$ represent for $^{27}_{13}$Al and $^{48}_{22}$Ti, respectively, which shows the magnitude of $R_{\mu e}$ is suppressed by $2-3$ orders compared to that of ${\rm Br}(\mu\rightarrow e\gamma)$. The right plot of Fig.~\ref{Fig5} shows the predictions of $R_{\mu e}$ in Ti as the functions of $x$ and ${\rm Br}(\tau\rightarrow\mu\gamma)$, where the values in the figure denote the branching ratio of ${\rm Br}(\tau\rightarrow \mu\gamma)$, the phase $\phi$ does not affect considerably and here $\phi=340^{\circ}$ is taken in Eq.~(\ref{Rmue1}). It is clear from Fig.~\ref{Fig5} that in a relatively large $\sin\theta_{13}\simeq{\cal O}(0.1)$, the upper bound of $R_{\mu e}$ requires the one of $\tau\rightarrow\mu\gamma$ should be less than a few $\times10^{-9}$ which is not more stringent than that of ${\rm Br}(\mu\rightarrow e\gamma)$. However, if the predicted ${\rm Br}(\mu\rightarrow e\gamma)$ is far below the planned $10^{-13}$ sensitivities, the $\mu-e$ conversion in nuclei can be a very competitive process to study LFV.

The CP asymmetry generated through the interference between tree and one-loop diagrams for the decay of the
heavy Majorana neutrino $N_{i}$ into $\eta$ and $(\nu,\ell_{\alpha})$ is given, for each lepton flavor $\alpha~(=e,\mu,\tau)$, by \cite{lepto2,Flavor}
 \begin{eqnarray}\nonumber
  \varepsilon^{\alpha}_{i}
  = \frac{1}{8\pi(\tilde{Y}^{\dag}_{\nu}\tilde{Y}_{\nu})_{ii}}\sum_{j\neq i}{\rm
  Im}\Big\{(\tilde{Y}^{\dag}_{\nu}\tilde{Y}_{\nu})_{ij}(\tilde{Y}_{\nu})^{\ast}_{\alpha i}(\tilde{Y}_{\nu})_{\alpha j}\Big\}g\Big(\frac{M^{2}_{j}}{M^{2}_{i}}\Big)~,
 \label{cpasym1}
 \end{eqnarray}
where the loop function $g(x)$ is given by
 \begin{eqnarray}
  g(x)&=& \sqrt{x}\Big[\frac{1}{1-x}+1-(1+x){\rm ln}\frac{1+x}{x}\Big]~.
  \label{decayfunction}
  \end{eqnarray}
Below temperature $T\sim M_{i}\lesssim10^{5}$ GeV, it is known that electron, muon and tau charged lepton Yukawa interactions are much faster than the Hubble expansion parameter rendering the $e$, $\mu$ and $\tau$ Yukawa couplings in equilibrium. Then, the CP asymmetries $\varepsilon^{\alpha}_{3}$ are approximately given\footnote{Due to ${\rm Im}[H_{3j}]=0$ for $\varphi_{1,2}=0$, the relation $\varepsilon^{\mu}_{3}+\varepsilon^{\tau}_{3}=-\varepsilon^{e}_{3}$ is satisfied if they are considered to the order of $x^{3}$.}, for $x\ll1$, by
 \begin{eqnarray}
  \label{normal1}
  \varepsilon^{e}_{3}&=& x^{2}\frac{\pi b}{2a}\frac{m_{3}}{m_{-}}(x\sin\phi+a\sin2\phi)~,\\
  \varepsilon^{\mu}_{3} &\simeq& x\frac{\pi b}{4a}\frac{m_{3}}{m_{-}}\{4a\sqrt{3}\cos\phi-x[\sqrt{3}+2a\cos\phi(\sqrt{3}\cos\phi+\sin\phi)]\}~,\nonumber\\
  \varepsilon^{\tau}_{3} &\simeq& x\frac{\pi b}{4a}\frac{m_{3}}{m_{-}}\{-4a\sqrt{3}\cos\phi+x[\sqrt{3}+2a\cos\phi(\sqrt{3}\cos\phi-\sin\phi)]\}~\nonumber
 \end{eqnarray}
where $g_{\nu}\simeq\sqrt{16\pi^{2}m_{3}/m_{-}}$ in Eq.~(\ref{mo1}) is used.
In these expressions, the values of the parameters $a,b,x,\phi$ are obviously determined from the analysis described in the previous section, whereas $m_{-}$ is arbitrary. However, as can be seen in Eq.~(\ref{taumu}) and the left plot of Fig.~\ref{Fig4}, the value of $m_{-}$ depends on the magnitude of $\bar{m}_{\eta}$ with a fixed value of ${\rm Br}(\tau\rightarrow\mu\gamma)$. In addition, generically, $|\varepsilon^{\alpha}_{3}|\gtrsim10^{-6-7}$ is needed to obtain a successful leptogenesis, which in turn indicates $m_{-}\simeq{\cal O}(100 {\rm eV})\lesssim$ a few keV.

\begin{figure}[t]
\begin{tabular}{cc}
\includegraphics[width=6.0cm]{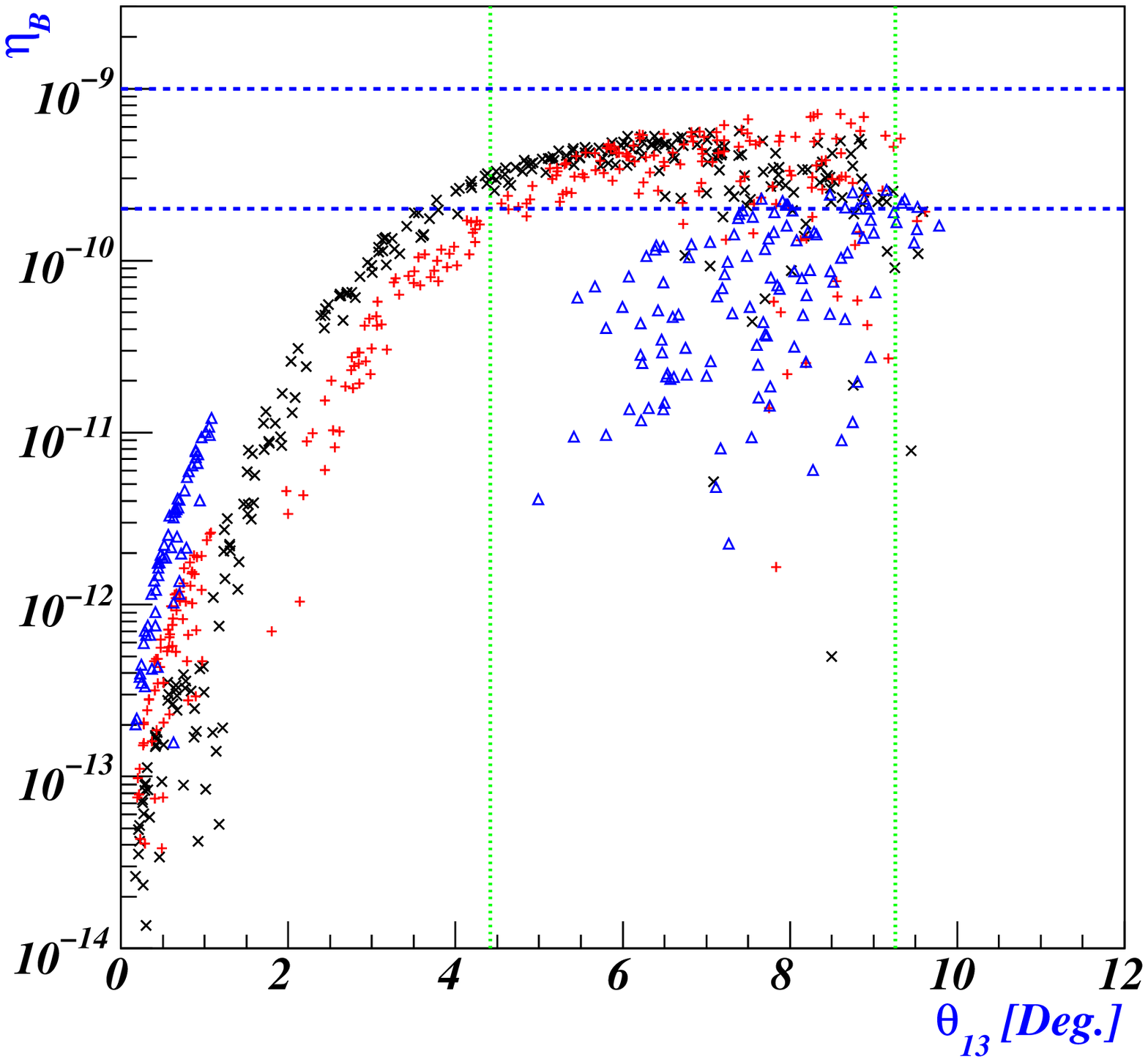}&
\includegraphics[width=6.0cm]{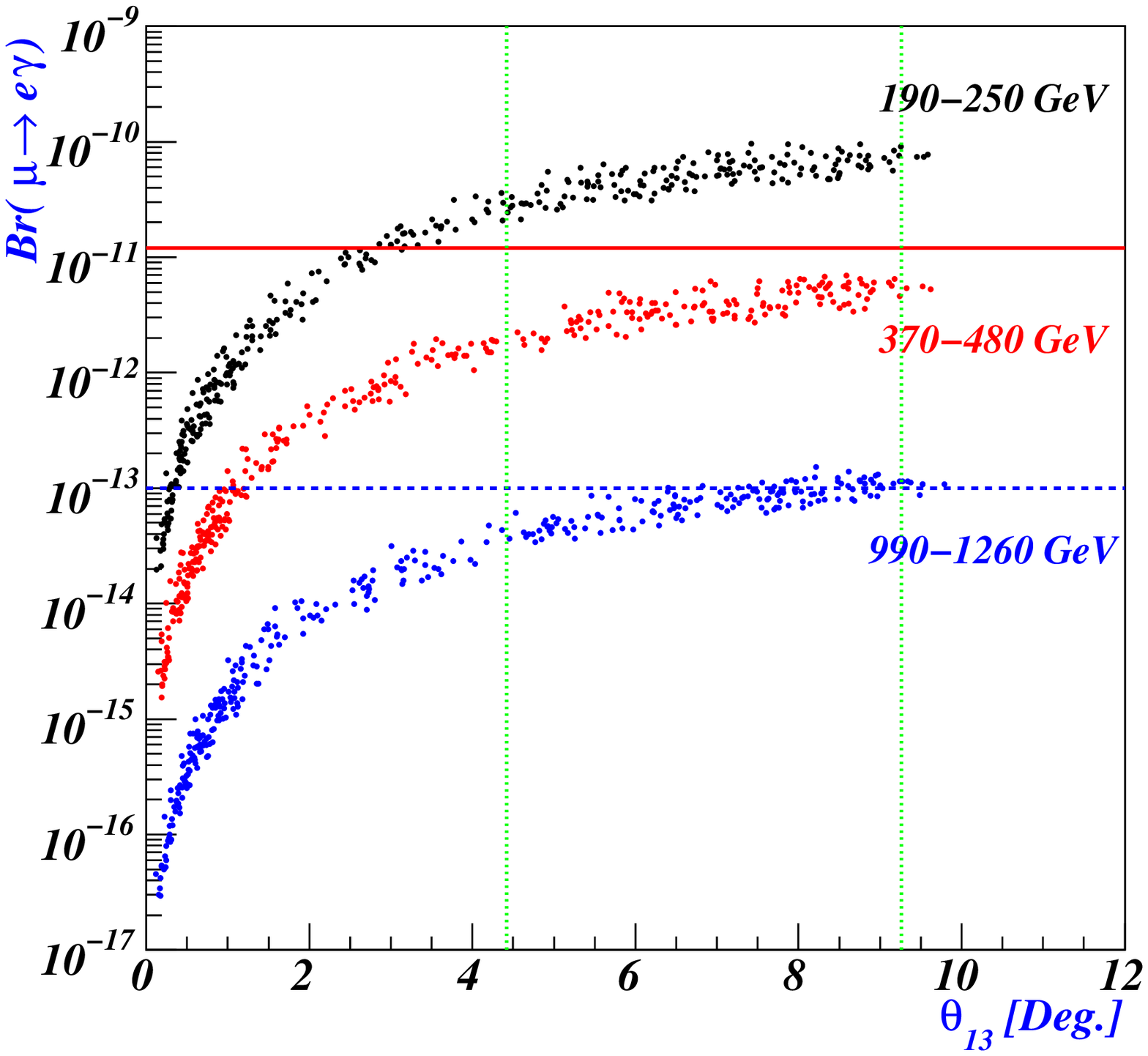}
\end{tabular}
\caption{\label{Fig6} (Left figure) Prediction of the baryon asymmetry $\eta_{B}$ as a function of the unknown mixing angle $\theta_{13}$, where the horizontal dotted lines correspond to the phenomenologically acceptable regions $2\times10^{-10}\leq\eta_{B}\leq10^{-9}$. (Right figure) Prediction of ${\rm Br}(\mu\rightarrow e\gamma)$ as a function of the unknown mixing angle $\theta_{13}$, where the values in the figure correspond to the mass of the lightest heavy Majorana neutrino, and the upper horizontal line and the below dotted-horizontal one correspond to the upper bound of ${\rm Br}(\mu\rightarrow e\gamma)<1.2\times10^{-11}, 10^{-13}$ for the present and for the forth coming future, respectively. In here $\delta_{N_3\eta}=5\times10^{-6}$ and $m_{-}=100$eV are taken, and the vertical dotted lines represent the experimental bounds in $1\sigma$ of the mixing angles $\theta_{13}$ in neutrino oscillations.}
\end{figure}

Once the initial values of $\varepsilon^{\alpha}_{i}$ are fixed, the final result of $\eta_{B}$ can be obtained by solving a set of flavor-dependent Boltzmann equations including the decay, inverse decay, and scattering processes as well as the nonperturbative sphaleron interaction.
Now, each lepton asymmetry for a single flavor in Eq.~(\ref{normal1}) is weighted differently by the corresponding wash-out parameter functioned by $K^{\alpha}_{i} =\Gamma(N_{i}\rightarrow \eta \ell_{\alpha})/H(M_{i})$ with the partial decay rate of the process $N_{i}\rightarrow\ell_{\alpha}+\eta$ and the Hubble parameter at temperature $T\simeq M_{i}$, and appears with different weight in the final formula for the baryon asymmetry\cite{Abada};
 \begin{widetext}
 \begin{eqnarray}
  \eta_{B}\simeq
  -2\times10^{-2}\Big\{\varepsilon^{e}_{3}\tilde{\kappa} \Big(\frac{151}{179}K^{e}_{3}\Big)
  +\varepsilon^{\mu}_{3}\tilde{\kappa}\Big(\frac{344}{537}K^{\mu}_{3}\Big)
  +\varepsilon^{\tau}_{3}\tilde{\kappa}\Big(\frac{344}{537}K^{\tau}_{3}\Big)\Big\}~.
  \label{etaB}
 \end{eqnarray}
 \end{widetext}
Here the wash-out factor $\tilde{\kappa}$ is given by
 \begin{eqnarray}
  \tilde{\kappa}\simeq\Big(\frac{8.25}{K^{\alpha}_{3}}+\Big(\frac{K^{\alpha}_{3}}{0.2}\Big)^{1.16}\Big)^{-1}~,
 \end{eqnarray}
where the wash-out parameters $K^{\alpha}_{3}$ associated with $N_{3}$ and the lepton flavors $\alpha=e,\mu,\tau$ are given as
 \begin{eqnarray}
  K^{e}_{3}&=& x^{2}\frac{8\pi^2}{3}\Big(\frac{m_{\ast}}{M_{3}}\Big)\Big(\frac{m_{3}}{m_{-}}\Big)\delta^{2}_{N_3\eta}~,\nonumber\\
  K^{\mu}_{3}&\simeq&8\pi^2\Big(\frac{m_{\ast}}{M_{3}}\Big)\Big(\frac{m_{3}}{m_{-}}\Big)\delta^{2}_{N_3\eta}(1+\frac{2x}{\sqrt{3}}\sin\phi)~,\nonumber\\
  K^{\tau}_{3}&\simeq&8\pi^2\Big(\frac{m_{\ast}}{M_{3}}\Big)\Big(\frac{m_{3}}{m_{-}}\Big)\delta^{2}_{N_3\eta}(1-\frac{2x}{\sqrt{3}}\sin\phi)~,
  \label{K-N}
  \end{eqnarray}
in which all $K$-factors are evaluated at temperature $T=M_{3}$, and $M_{3}$ is the lightest of the heavy Majorana neutrinos.
Here $m_{\ast}=\big(\frac{45}{2^{8}\pi^{5}g_{\ast}}\big)^{\frac{1}{2}}M_{\rm Pl}\simeq2.83\times10^{16}$ GeV with the Planck mass $M_{\rm Pl}=1.22\times10^{19}$ GeV and the effective number of degrees of freedom $g_{\ast}\simeq g_{\ast \rm SM}=106.75$, and the degeneracy between $M^{2}_{3}$ and $\bar{m}^{2}_{\eta}$ is given by~\cite{Ahn:2010cc,Gu:2008yk}
 \begin{eqnarray}
  \delta_{N_3\eta}\equiv1-\frac{\bar{m}^{2}_{\eta}}{M^{2}_{3}}~,
  \label{degeneracy}
 \end{eqnarray}
in which $\bar{m}^{2}_{\eta}\simeq M^{2}_{3}$ is necessary for enormously huge wash-out factors to be tolerated. Note here that wash-out factors associated with $N_{1,2}$ and the lepton flavors $\alpha=e,\mu,\tau$ are enormously huge compared to the factors $K^{e,\mu,\tau}_{3}$, and therefore the generated lepton asymmetries associated with $N_{1,2}$ are strongly washed out due to $K^{\alpha}_{i} \simeq m_{\ast}g^{2}_{\nu}\delta^{2}_{N_i\eta}|\tilde{Y}_{\nu\alpha i}|^{2}/M_{i}|\tilde{Y}_{\nu}|^{2}_{ii}$ where for $M^{2}_{i}\gg\bar{m}^{2}_{\eta}$ the degree of degeneracy $\delta_{N_i\eta}$ is going to be 1. It is clear that, if the value $1/m_{-}M_{3}$ is constrained by both low energy neutrino data, LHC signal and LFV constraints with $\bar{m}_{\eta}\simeq M_{3}=bM$, the wash-out factors $K^{e}_{3}$ and $K^{\mu\tau}_{3}$ are only dependent on the parameter $\delta_{N_3\eta}$ which makes the difference to the flavor effects. As can be seen in Eqs.~(\ref{normal1},\ref{K-N}), since the lepton asymmetries in $\mu$ and $\tau$ flavors are equal but opposite in sign to the first order, i.e. $\varepsilon^{\mu}_{3}\approx-\varepsilon^{\tau}_{3}$, satisfying $\varepsilon^{\mu}_{3}+\varepsilon^{\tau}_{3}=-\varepsilon^{e}_{3}$, and the wash-out parameters in $\mu$ and $\tau$ are almost equal $K^{\mu}_{3}\approx K^{\tau}_{3}\gg K^{e}_{3}$, the effects of wash-out factor related with $N_{3}$ can play a crucial role in a successful leptogenesis according to the size of $\delta_{N_3\eta}$. In our scenario, although $\delta_{N_3\eta}$ does not much affect the results for low energy neutrino observables obtained in sec. II, the predictions of the baryon asymmetry $\eta_{B}$ strongly depends on the quantity $\delta_{N_3\eta}$ due to the size of wash-out parameters.
The left figure in Fig.~\ref{Fig6} shows the prediction of the baryon asymmetry for $\delta_{N_3\eta}=5\times10^{-6}$, which indicates a successful leptogenesis favors a relatively large value of $\theta_{13}$; flavor effects are shown according to the varying scale of heavy neutrino $N_{3}$ for $190-250$ GeV (red-crosses), $370-480$ GeV (black-multiplies) and $990-1260$ GeV (blue-triangles).

Let us consider possible implications for neutrino parameters from future LFV searches with the analysis obtained in Eqs.~(\ref{Bab},\ref{LFV1}).
Using Eq.~(\ref{Bab}), the constraints of Eq.~(\ref{exp}) can be replaced by
 \begin{eqnarray}
  |B_{\tau\mu}|^{2}&<& 5.6\times 10^{-14}\times r^{4} \left({\bar{m}_\eta\over r~{\rm GeV}}\right)^4,\nonumber\\
  |B_{\tau e}|^{2}&<& 1.5\times 10^{-13}\times r^{4} \left({\bar{m}_\eta\over r~{\rm GeV}}\right)^4,\nonumber\\
  |B_{\mu e}|^{2} &<& 1.5\times 10^{-17}\times r^{4} \left({\bar{m}_\eta\over r~{\rm GeV}}\right)^4.
 \label{LFVBound}
 \end{eqnarray}
where the parameter $r$ GeV is the mass of lightest heavy Majorana neutrino $N_{3}$ for $M_{3}\simeq\bar{m}_{\eta}$, which indicates if both $m_{-}$ (which can be constrained by both a successful leptogenesis in our scenario and $g_{\nu}\lesssim1$ as $8{\rm eV}\lesssim m_{-}\simeq{\cal O}(100{\rm eV})$) and $\theta_{13}$ are given these constraints depend on the mass of DM. Since, numerically, for $m_{-}=100$ eV and $x=0.2, \phi=340^{\circ}$ the value of $|B_{\tau\mu}|^{2}$ is given around $5\times10^{-6}$, from the constraint $\tau\rightarrow\mu\gamma$ in Eq.~(\ref{LFVBound}) the mass of $\bar{m}_{\eta}$ should be lie above $100$ GeV at least. However, from Eqs.~(\ref{Bab},\ref{LFVBound}) it is clear that the most of constraints comes from ${\rm Br}(\mu\rightarrow e\gamma)$ which is sensitive to the value of $|U_{e3}|$, $m_{-}$ and the mass of DM. Since the branching ratio of $\ell_\alpha\rightarrow\ell_\beta\gamma$ explicitly depends on the $m_{\pm}$ and the unknown mixing angle $\theta_{13}$, in near future searching $\theta_{13}, m_{-}$ and ${\rm Br}(\ell_\alpha\rightarrow\ell_\beta\gamma)$ indicates finding the mass of DM. For example, considering the experimental bounds both of $\theta_{13}$ in $1\sigma$ in which the value of baryon asymmetry is well explained and of the branching ratio $\mu\rightarrow e\gamma$, for $x=0.2, \phi=340^{\circ}$ and $m_{-}=100$ eV we see that from Eq.~(\ref{LFVBound}) $|B_{\mu e}|^{2}\simeq10^{-7}$ the mass of DM should be larger than $285$ GeV corresponding to the constraint of $\mu\rightarrow e\gamma$. As stressed above, given the DM mass and the missing energy, $\theta_{13}$ is crucial for the prediction of $\mu\rightarrow e\gamma$ branching ratio. To see its dependence, we plot the LFV prediction of $\mu\rightarrow e\gamma$ as the unknown mixing angle $\theta_{13}$ with $m_{-}=100$ eV and varying DM masses, in Fig.~\ref{Fig6}. Since the below horizontal line in the right figure in Fig.~\ref{Fig6} indicates the present upper bound of ${\rm Br}(\mu\rightarrow e\gamma)$, there is a lower bound, $M_{3}\gtrsim285$ GeV, of the mass of DM with both the experimental bound of $\theta_{13}$ in $1\sigma$ and the value of $\eta_{B}$ in thermal leptogenesis scenario. MEG experiment searching $\mu\rightarrow e\gamma$ is expected to reach ${\rm Br}(\mu\rightarrow e\gamma)\sim{\cal}(10^{-13}-10^{-14})$~\cite{Mori:2007zza}, which reads the right figure in Fig.~\ref{Fig6} with~Eq.~(\ref{LFV1}).
Therefore, in near future we can see the mass of DM indirectly.
From Fig.~\ref{Fig6}, with a missing energy $m_{-}=100$ eV the requirement for the successful leptogenesis consistent with $\sin\theta_{13}\simeq{\cal O}(0.1)$ can be compatible with the existing constraint on ${\rm Br}(\mu\rightarrow e\gamma)$ if  $\bar{m}_{\eta}\gtrsim285$ GeV.

\section{conclusion}

We have investigated the relation between neutrino parameters and LFV prediction, in the light of the deviation of TBM and the recent precision oscillation data. Particularly, we have examined how LFV is related to the reactor angle $\theta_{13}$, LHC/DM signal with a successful leptogenesis in a radiative seesaw model through $A_4$ flavor symmetry breaking which reads the deviations from TBM mixing angles. We have focused on the normal mass spectrum of light neutrino giving a successful leptogenesis at electroweak (EW) scale or more, where a successful leptogenesis requires generically lepton asymmetry $|\varepsilon^{\alpha}_{i}|\gtrsim10^{-6-7}$ which in turn indicates LHC/DM signal $m_{-}\simeq{\cal O}$(100eV)$\lesssim$ a few keV in our scenario, and we have shown that the leptogenesis scale can be determined by the branching ratio of $\tau\rightarrow\mu\gamma$ if the missing energy $m_-$ defined in Eq.~(\ref{mo1}) is measured. Since the ratio of the branching ratios for ${\rm Br}(\mu\rightarrow e\gamma)/{\rm Br}(\tau\rightarrow\mu\gamma)$ is strongly dependent on the value of $\theta_{13}$ in our scenario, if future experiments of neutrino and LFV would measure the nonvanishing $U_{e3}$ and ${\rm Br}(\mu\rightarrow e\gamma)$ respectively, the branching ratio of $\tau\rightarrow\mu\gamma$ can be predicted with a successful leptogenesis, and which in turn indicates the mass of DM ($\simeq$ leptogenesis scale) and the missing energy can be strongly bounded by ${\rm Br}(\tau\rightarrow\mu\gamma)$.
In addition, we show the magnitude of $\mu-e$ conversion $R_{\mu e}$ in Ti is suppressed by $2-3$ orders compared to that of ${\rm Br}(\mu\rightarrow e\gamma)$.
Finally, we have shown that, for example, with a missing energy $m_{-}=100$ eV the requirement for the successful leptogenesis consistent with $\sin\theta_{13}\simeq{\cal O}(0.1)$ can be compatible with the existing constraint on ${\rm Br}(\mu\rightarrow e\gamma)$ if the mass of DM $\bar{m}_{\eta}\gtrsim285$ GeV.

\acknowledgments{
YHA is supported by the National Science Council of R.O.C. under Grants No:
 NSC-97-2112-M-001-004-MY3.}


\end{document}